# OUT-OF-PLANE CMOS COMPATIBLE MAGNETOMETERS

*Mehdi El Ghorba[1,3], Nicolas André[2], Stanislas Sobieski[1], Jean-Pierre Raskin[1]*

[1]Microwave Laboratory, [2]Microelectronics Laboratory
Université catholique de Louvain, Louvain-la-Neuve, BELGIUM
(Tel: +32 10 47 23 09, E-mail: raskin@emic.ucl.ac.be)
[3]Master in Micro and Nanotechnologies for IC Technology
INPG, Grenoble, FRANCE; EPFL, Lausanne, SWITZERLAND;
Politecnico di Torino, Torino, ITALY
(Tel: +32 10 47 25 66-mail: mehdi.el-ghorba@polymtl.ca)

## Abstract

Three-dimensional MEMS magnetometers with use of residual stresses in thin multilayers cantilevers are presented. Half-loop cantilevers based on Lorentz-force deflection convert magnetic flux in $\Delta V$ changes, thanks to piezoresistive transducers mounted in Wheatstone bridge. Magnetic field in the order of 10 Gauss was measured with a sensitivity of 0.015 mV/Gauss. A Finite Element Model of the device has been developed with Ansys for static and dynamic simulations. Novel out-of-plane ferromagnetic nickel plate magnetometer is also presented.

## 1. INTRODUCTION

During last decade many – CMOS compatible MEMS magnetic sensors have been developed [1-3], all of them were optimized for in-plane magnetic field measurement. In this work an out-of-plane CMOS compatible magnetic field sensor is presented. Out-of-plane sensing was achieved by using the self-assembled three-dimensional (3-D) MEMS process described in [4]. This simple and reliable process uses residual stress originated from thermal annealing for the assembling of multilayered structures.

## 2. DESCRIPTION

### 2.1 Sensing principle

For our first type, the out-of-plane magnetic flux is converted into a mechanical force on the M-shaped cantilever by Lorentz force $F$ (Fig. 1-a and 1-b):

$$F = I \cdot L \cdot B \cdot \sin(\theta) \quad (1)$$

where $I$ is the half-loop current, $L$ is the top beam length, $B$ is the magnetic field across this beam and $\theta$ is the angle between the magnetic field and the current flowing into the top beam. Maximal stress at the anchor of beams can be approximated by:

$$\sigma_{max} = \frac{6l}{wt^2}\frac{F}{3} \quad (2)$$

where $l$ is the beam length, $w$ the width and $t$ the thickness. A piezoresistive gauge converts the beam bending into an electrical signal by use of a Wheatstone bridge. Considering only the axial stress, piezoresistors variation $\Delta R$ is given by:

$$\frac{\Delta R}{R} = 4\frac{\Delta V_{out}}{V_{bias}} = \pi_l \sigma \quad (3)$$

where $\pi_l$ is the longitudinal piezo coefficients, $\Delta V_{out}$ the bridge output and $V_{bias}$ the bias voltage. The $\pi_l$ coefficient depends mainly on process parameters like crystallographic orientation and doping level, as well as the working temperature.

For our second type, a magnetized nickel monolayer aligns to the direction of the applied external magnetic flux. This resulting torque can be described by the following:

$$T = \overline{M} \times \overline{B} \quad (4)$$

where $M$ is the magnetization of the nickel monolayer. The plate is suspended by 2 cantilevers (Fig. 1-c.). Piezoresistors at the anchors allow to extract the field intensity the same way as described above.

### 2.2 Noise modelling

Flicker ($1/f$) and thermal (electrical - mechanical) noise are the two dominant internal noise sources of a piezoresistive beam [5]. The thermal power noise spectral density for a resistance $R$ and a mechanical damping $D$ at the temperature $T$ are given by:

$$S_{T-E} = 4k_b TR \qquad S_{T-M} = 4k_b TD \quad (5)$$

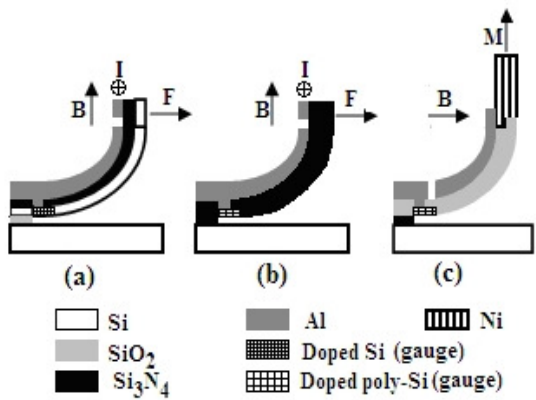

*Fig. 1. Cross-section view showing the operation of the magnetometers, where B is the applied magnetic field, I the current in the M-shaped loop, M the magnetization and F is the applied force on the structure. (a) SOI–Lorentz (out-of-plane field sensing), (b) Si-Lorentz (out-of-plane plane sensing), (c) Si-ferromagnetic (in-plane sensing).*

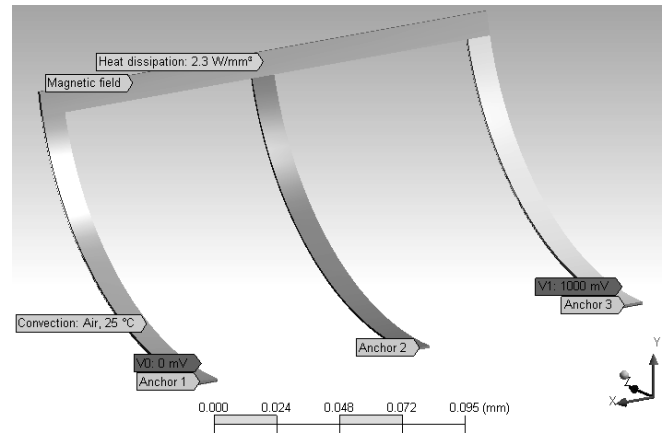

*Fig 2. Ansys finite element model with buondaries condition and external forces.*

where $k_b$ is the Boltzmann constant. For this kind of structure the mechanical noise can be neglected in comparison with the electrical one [2, 5]. Flicker power noise spectral density is given by:

$$S_f = \frac{\alpha V^2}{lwt\rho_0}\frac{1}{f} \quad (6)$$

where $l$, $w$ and $t$ are the gauge dimensions, $V$ is the biased voltage across a resistor, $f$ is the frequency, and $\alpha$ is a dimension independent device parameter which is between $2\times10^{-6}$ and $6\times10^{-6}$ in single crystal silicon. Flicker noise dominates below 100 Hz and can be neglected above this frequency [5].

### 3. DESIGN AND FABRICATION

#### 3.1 Ansys Finite Element Model

A finite elemet model was developed with Ansys Multiphysics (Fig. 2), tacking into acounount the mechanical lorentz forces, the thermal dissipation by joule effect, air damping and Boundaries conditions. This model allowed us to understand the static and dynamic behaviour.

#### 3.2 Fabrication

The device was fabricated in both Silicon-on-Insulator (SOI) (Fig. 1-a) and Si-Bulk (Fig. 1-b and 1-c) technologies with few differences between them. In SOI technology, the trilayered structure is composed of the active thin Si layer (i.e. top SOI layer ≈ 100 nm), 280 nm-thick $Si_3N_4$ and 1 μm-thick Al. The stoechiometric nitride layer is obtained by LPCVD deposition at 800°C and the pure aluminum layer is evaporated in an E-gun vacuum system at 150°C. Piezoresistive gauges are made from the top monocristalline silicon, N-doped by phosphorus solid-source diffusion technique. Buried oxide of 400 nm-thick is used as a sacrificial layer and is removed in HF-based solution.

In bulk-Si technology, beams are composed only by a $Si_3N_4$ (350 nm) – Al (1 μm) bi-layer with the same deposition techniques. To replace the sacrificial BOX layer, a 1 μm-thick PECVD oxide layer is initially added on blank wafer. The piezoresistive material must be deposited and is made by LPCVD polysilicon obtained at 625°C and consecutively N-doped (Fig. 1-b and 4.). The release solution is HF-based.

With almost the same process we also developed a ferromagnetic sensor (Fig. 1-c and 4). The major difference is that a thin rectangular layer (500 nm) of nickel is deposited at the free end of two cantilevers and then magnetized. Because of the presence of nickel, HF-based release is prohibited. Instead, $SF_6$ plasma is used successfully as isotropic dry etchant. With an appropriate beam length, the rectangular magnetized Ni layer is perpendicular to the surface (monolayer without gradient of stress).

#### 3.3 Thermal Annealing

For both cases, the assembly of these 3-D microstructures relies on the control of the residual internal stress in thin layers due to thermal expansion misfit. After release the last step of the process consists in a thermal annealing in order to promote the assembling of the microstructures by modifying the stress state in the upper plastic layer (Al) [4, 6]. This thermal annealing treatment corresponds to a forming gas $H_2/N_2$ at 430°C for 30 minutes. This thermal treatment is similar to the last step of our SOI-CMOS process used to improve the metal-polysilicon contacts as well as to reduce the gate oxide interface traps. By increasing temperature, the stress in Al layer becomes heavily compressive, reaches the yield stress to finally lead to a fully plastic layer.

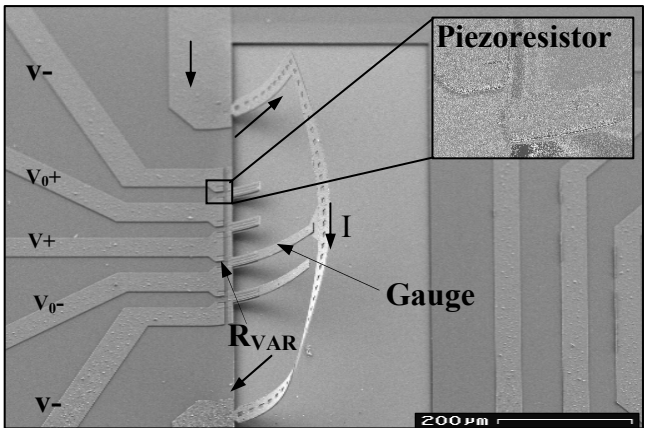

*Fig. 3. SEM picture of a CMOS compatible magnetometer (Si-Lorentz) with its integrated Wheatstone bridge.*

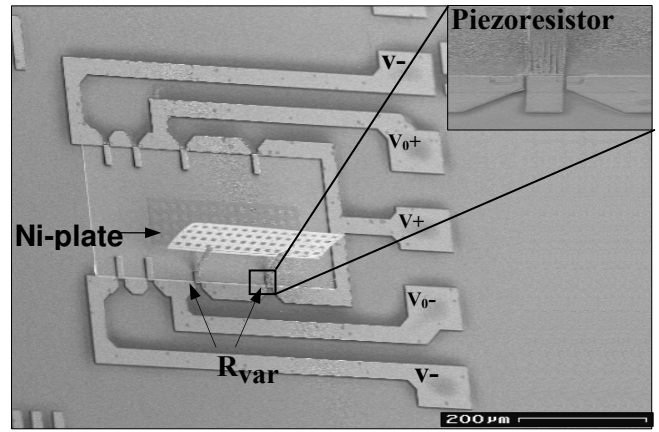

*Fig. 4. SEM picture of a magnetometer (Si-ferromagnetic) with its integrated Wheatstone bridge.*

Upon cooling, stress in aluminium becomes more tensile than after deposition, helping the microstructure to lift up. This stress increase results from the plasticity of aluminium layer, and is affected also by the hardening capacity and temperature dependence of the flow properties. Stress value in the aluminium film required to obtain self-assembled structures is above 150 MPa. Such high stress values require annealing temperatures above 400°C. Measured stresses in aluminium films after 30 s RTA are around 150 MPa at 400°C. A finite element model which agrees quite well with experimental deflections was developed in ABAQUS to simulate this thermo-mechanical process [4].

The gauge is electrically and thermally isolated with $Si_3N_4$ layer from the half-loop current to avoid detrimental effects caused by the main current *I* (which serves to deflect the beam as shown in Eq. 1). The system can work under DC or AC supply, however for bigger deflections and less noise it must operate at its resonant frequency and moreover by applying a square alternative current - thermal expansion by Joule effect of the main current remains constant. The Wheatstone bridge is integrated into the device; its static resistors are also placed at the fixed end of deflected beams in order to have the same resistance as the variable one at rest. A full representation of the device is shown in Fig. 3 and 4.

## 4. RESULTS AND DISCUSSION

In a monolithic chip, Wheatstone bridge output voltage is used as the input to a signal processing circuit; however, this advanced step of integration has not been achieved in this work. Tests were performed by varying magnetic field strength of an electromagnet as well as the half-loop current for devices with different geometries.

As shown in Fig. 5, the response is linear in a high range of the magnetic field (from 0.1 mTesla to a few tens of mTeslas) with a small offset at 0 V. It varies quadriticaly with the loop current and is mainly due to the thermal dissipation by Joule effect. This offset is about 0.03 mV (0.02 mT) for a half-loop current of 10 mA and 0.1 mV (0.10 mT) for a current of 50 mA. In AC mode we used rectangular alternative current to keep the thermal dissipated energy constant. Thermal dissipation by Joule effect increases the thermal noise and limits the current which can be applied and then the sensitivity, it also decreases the Signal-to-Noise Ratio. However, it does not affect the piezoresistive gauge, which is highly temperature dependent, since it is isolated in the middle beam (Fig. 1 and 3). In our case thermal effects can be neglected for currents smaller than mAmps.

Built with a similar process as described in [7] a 3-D inductance (Fig. 6-b) deflects if a current go through and in presence of a magnetic field as described by Eq. 1. By using an AC current through the inductance, and sweeping the frequency, the resonant frequency has been found. Fig. 6-a shows the inductance behavior at the resonant frequency. For the Lorentz-force based sensor shown in Fig. 3, a damping can be observed preventing the resonance of the device probably due to the central beam used to extract the magnetic field intensity.

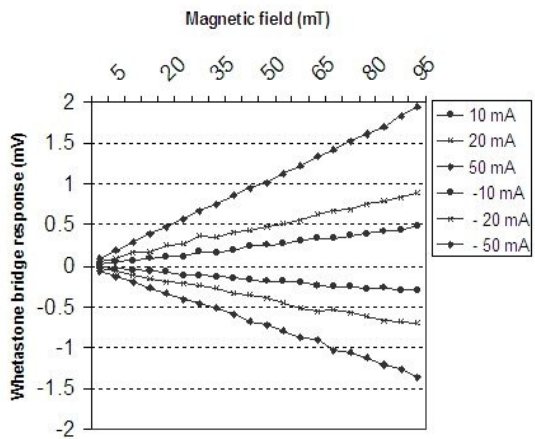

*Fig. 5. Sensor response in mV for different rectangular currents at 4 kHz.*

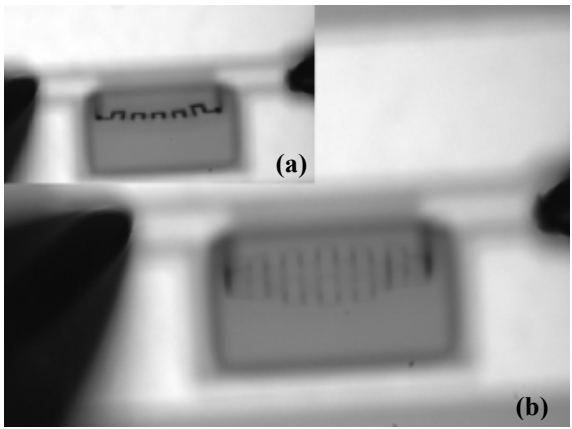

*Fig. 7. Top view of a magnetometer (a) at rest and (b) under ±0.75 V at 4 kHz and a magnetic field of 0.1 T*

For the ferromagnetic sensor shown in Fig. 3 a deflection of more than 10 µm has been observed for a field of 0.4 T perpendicular to the direction shown in Fig. 1-c. A torque is applied on the magnetized nickel plate since the released layer is not exactly aligned with the magnetic field applied. The deflection observed is similar to the one measured for the Lorentz-force based sensor (Fig. 3), thus similar sensitivity is expected.

## 5. CONCLUSION

An out-of-plane Lorentz magnetometer and an in-plane ferromagnetic magnetometer have been developed with few additional steps to a standard CMOS process. These devices are perpendicularly to the chip plane by using the residual stress in multilayer beams. Integrated with two perpendiculars in plane devices, it is possible to achieve a three-dimensional magnetic field sensing. A better resolution can be achieved by optimising the design and the process.

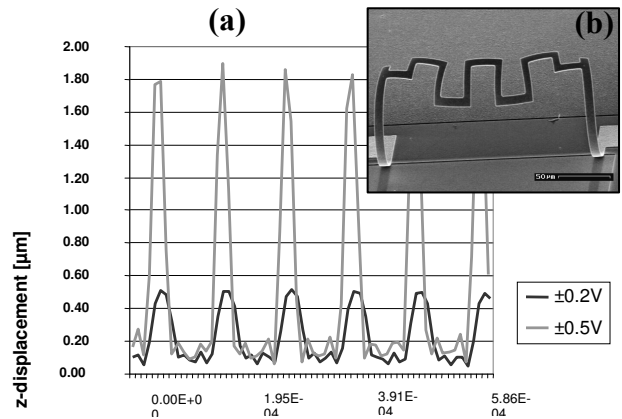

*Fig. 6. (a) Time domain response under a magnetic field of 0.4 mT (±0.5 V, ±0.2 V - 6 kHz) (b) SEM picture of an inductance at rest described in [7].*